# A MULTI-AGENT SYSTEM APPROACH IN EVALUATING HUMAN SPATIO-TEMPORAL VULNERABILITY TO SEISMIC RISK USING SOCIAL ATTACHMENT

JULIUS BAÑGATE[1,2,3], JULIE DUGDALE[1,3], ELISE BECK[2,3] & CAROLE ADAM[1,3]
[1]Laboratoire d'Informatique de Grenoble (LIG)
[2]Politiques publiques, Action politique, Territoires (PACTE)
[3]University Grenoble-Alps, Grenoble, FRANCE

ABSTRACT

Social attachment theory states that individuals seek the proximity of attachment figures (e.g. family members, friends, colleagues, familiar places or objects) when faced with threat. During disasters, this means that family members may seek each other before evacuating, gather personal property before heading to familiar exits and places, or follow groups/crowds, etc. This hard-wired human tendency should be considered in the assessment of risk and the creation of disaster management plans. Doing so may result in more realistic evacuation procedures and may minimise the number of casualties and injuries. In this context, a dynamic spatio-temporal analysis of seismic risk is presented using SOLACE, a multi-agent model of pedestrian behaviour based on social attachment theory implemented using the Belief-Desire-Intention approach. The model focuses on the influence of human, social, physical and temporal factors on successful evacuation. Human factors considered include perception and mobility defined by age. Social factors are defined by attachment bonds, social groups, population distribution, and cultural norms. Physical factors refer to the location of the epicentre of the earthquake, spatial distribution/layout and attributes of environmental objects such as buildings, roads, barriers (cars), placement of safe areas, evacuation routes, and the resulting debris/damage from the earthquake. Experiments tested the influence of time of the day, presence of disabled persons and earthquake intensity. Initial results show that factors that influence arrivals in safe areas include (a) human factors (age, disability, speed), (b) pre-evacuation behaviours, (c) perception distance (social attachment, time of day), (d) social interaction during evacuation, and (e) physical and spatial aspects, such as limitations imposed by debris (damage), and the distance to safe areas. To validate the results, scenarios will be designed with stakeholders, who will also take part in the definition of a serious game. The recommendation of this research is that both social and physical aspects should be considered when defining vulnerability in the analysis of risk.

## 1 INTRODUCTION

Human vulnerability during disasters is subject to the complex confluence of dynamic individual, social, spatial and temporal elements. The mobility of people in space and time is influenced by societal structure, norms, and individual capacities (human factors). Social attachment theory describes the social nature of human behaviour during disasters and states that individuals will seek the proximity of attachment figures [1]. This influences the choices and actions that individuals make during evacuations. This paper describes how social attachment has been implemented in a multi-agent model using a belief-desire-intention (BDI) approach. The model, called SOLACE for SOciaL Attachment & Crisis Evacuations, was coded using the GAMA platform [2]. SOLACE aims to provide a realistic model of human behaviour during seismic crisis. The novelty of the model is the integration of social attachment and GIS data. This paper presents initial results evaluating different spatial and temporal scenarios influencing human vulnerability to seismic risk when applied to two districts of the city of Grenoble in the French Alps.

This paper is organised as follows: Section 2 presents related work. Section 3 provides a brief description of the model. Section 4 details the experiments. Section 5 describes initial results. Section 6 discusses the results, presents the conclusions and describes future work.

## 2 RELATED WORK

Research on seismic risk and vulnerability has largely focused on physical aspects such as damage to structures and the prediction/forecasting of future events. Seismic crisis management plans in general are top-down institutional responses (command and control) and are geared towards logistics and supply chain. Evacuation plans for seismic events are based on expected, rather than actual, human behaviours and lack realism.

The need for more research on human behaviour during crisis (moderate earthquakes in particular) has been noted in several studies [3][4]. In order to prepare and respond to largely unpredictable events it is essential that more is known about how people really behave in a crisis situation. In addition existing earthquake models (maps, algorithms) need calibration and robust testing. Furthermore the structural, geological, and geotechnical data that are needed by models lack detail, completeness and coverage. These facts illustrate the big gap in the required knowledge for developing preparedness and resilience.

Social theories on human behaviour during disasters are presented in [1], [5]. Mawson's social attachment theory posits that when individuals are under threat they seek the proximity of attachment figures. Attachment figures include family members, friends, pets, colleagues, authorities and even strangers. Familiar objects, places, and information present within the sphere of social interaction are also considered as attachment figures. During a threat, affiliation to attachment figures is activated and can lead to pre-evacuation behaviours such as seeking family members, milling, herding, protecting property, seeking pets, helping others/strangers, etc. Such attachments can influence an individual's decisions, destinations (goals), social interactions, and speed/direction of movement.

Humans fundamentally desire and maintain interpersonal attachments [6]. However the strength of bonds varies depending on the attachment figure, e.g. a person may be more strongly attached to their mother than their pet. The strength of emotional bonds for different relationships (partner, parents, kin, friends, acquaintance, strangers) in a social network in different European countries has been quantified in Suvilehto, et al [7]. Touch (in different bodily areas) was used to quantify bond strength. Social situations that involve touch include greeting, parting, giving attention, helping, consoling, calming, and giving pleasure. It should be noted that variations in bond strength and in ordering/priority may be due to cultural differences. Table 1 shows some results for France.

Table 1: Mean strength of emotional bonds for France derived from Suviletho et al, 2015

| Relation | Partner | Parent | Sibling | Kin | Friend | Acquaintance | Stranger |
|----------|---------|--------|---------|------|--------|--------------|----------|
| Bond     | 8.82    | 7.77   | 7.51    | 5.29 | 7.57   | 3.84         | 2.17     |

Familiarity and bond strength mean that we are more able to perceive and recognise attachment figures from a distance, within a group or in a complex environment. The perception distance refers to what one sees (facial expression, body language, posture, gesture, actions, signals/signs), hears (voice, sound, calling by name, call for help, screams, warnings) [8] [9] and cognitive representations (knowledge, memory). The perception distance can differ for each context and varies for each individual. This can be due to psychosocial bias (due to attachment, cognitive capacity, emotions), sensory limitations (due to age, disability), and/or environmental constraints such as location/proximity, occlusion due to barriers, diminished signals due to physical laws (speed of light/sound), or environmental conditions (foggy, night, blackout).

The perception of shaking intensity during an earthquake depends on the ground motion and the building's response [10]. Also, [10] found that there is a relationship between peak seismic intensities and human behaviour. Human responses to different earthquake shaking intensities have been reported in [11] [12] [13]. Similar to the findings of [11], variations in resulting behaviours in different intensities have also been observed in [14], [15], and [16]. The psychological and social responses to disasters in general are reported by [17].

Spatial configuration of the built environment limits movement, especially after a disaster [16] [18] 19], and the visual perception range of pedestrians to open spaces (parks or wide streets) that constitute safe areas following an earthquake. The urban morphology channels pedestrian movement and facilitates social interaction. The volume of people in public areas changes, depending on the economic and social activities during different times of the day or week (urban pulse) [20]. Barriers to movement include fixed structures (buildings or fences), natural features (rivers/streams or bodies of water), and dynamic objects (cars or other people). Debris from damaged structures during disasters also block normal passageways [21] [4]. The increasing population in cities presents a large design challenge to planners and policy makers in building safe and disaster resilient cities.

Agent based modelling (ABM) provides the opportunity to develop models of crisis situations at the micro-, meso- and macro- scale, and to test them using computer simulation. A trend in ABM is the integration of psychological and social aspects of human behaviour into agents in order to create more realistic models. The goal is to enhance current models with agents that are social, have the capability to decide, can identify and associate themselves with a group. Previous models have integrated some attachment figures for crisis evacuations, e.g. EPES, ESCAPES and EXITUS. A complete review of previous models is given in [5].

Finally, in the field of ABM, the BDI approach is useful to reproduce human behaviour [22]. It attempts to capture how humans reason with beliefs (internal knowledge), desires (what it likes to achieve), and intentions (actions, guided by a library of plans).

## 3 DESCRIPTION OF THE MODEL

### 3.1 Multi-disciplinary approach

The SOLACE model has been developed by drawing upon social, computer and geospatial sciences (Fig. 1). Earthquake crisis evacuation behaviours are gathered from reports, video-recordings and literature. Social concepts and theories on human behaviour during crises are used as a lens for analysing the data, developing the conceptual model, validating the results with stakeholders, and the use of serious games. Formalization of the agent-based model using BDI with GAMA used a variety of computer science techniques. In particular the realistic spatial context of the simulated environment includes integrating different datasets concerning to urban layout, seismic risk and vulnerability, mobility, demography, etc. This spatial context was developed using geospatial data processing workflows in GIS.

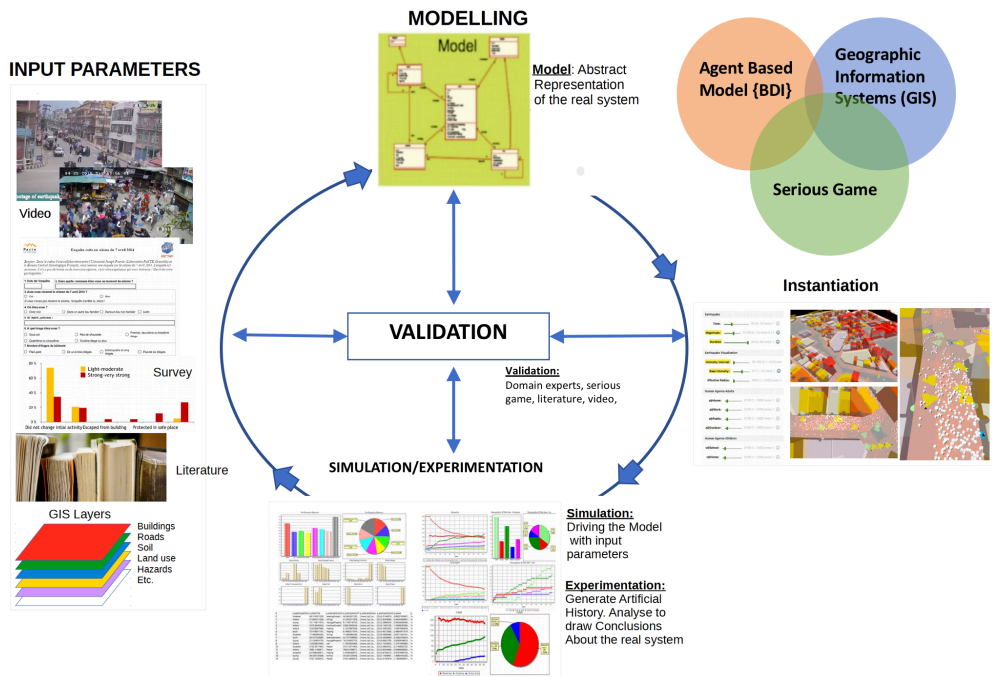

Figure 1: Multidisciplinary methodology

### 3.2 Social model

Social attachment is the social theory that has been implemented in SOLACE and which drives an individual's goals and social behaviours. This is expressed as bond strength, social distance, and perception distance. Following Fig. 2 a social distance from 1 to 10 is used to set the priority in perceiving attachment figures from a reference adult/individual. For example for an adult, a priority of 1 is given to their child; priority 2 to their partner, and 3 to their parents (from the data in table 1). However these priorities may be changed in the simulator interface. Since data from [7] did not include children it has been assumed that they have a higher priority than a partner. Note that the social distance to oneself is 0. Details of these priority rankings are discussed in [2]. Adopting usual social norms for relationships, the attachment bond strength is strongest for the nearest social distance; as is the case for an altruistic scenario. In an egoistic scenario, the social bond strength is 0 for oneself and there is no social bond to other agents; there is no social interaction. Thus egoistical agents will only perceive themselves, give themselves priority, and try to protect and preserve only themselves.

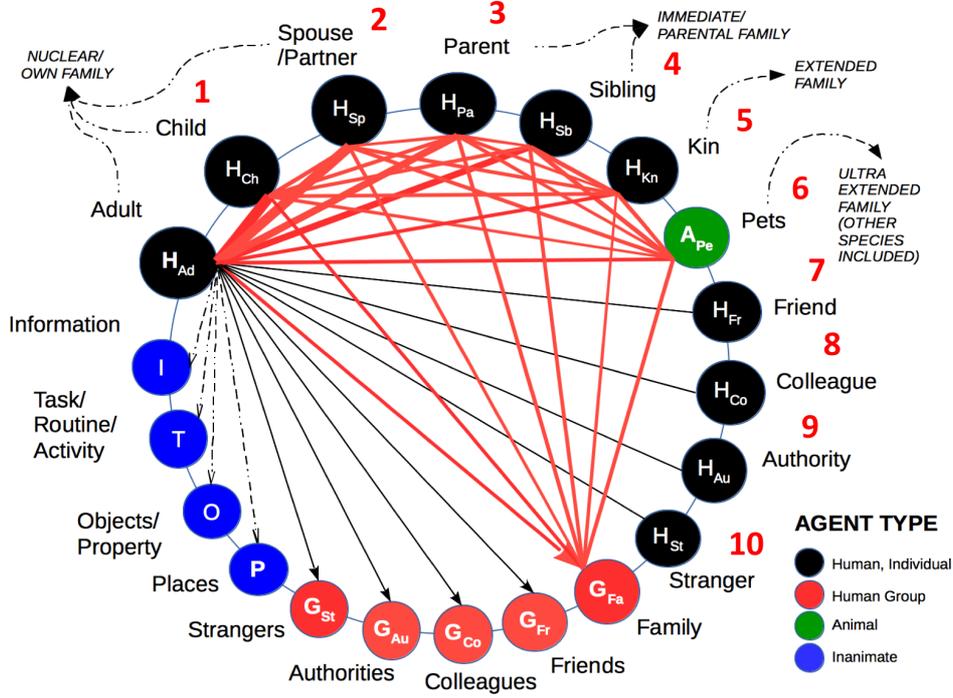

Figure 2: Social attachment network

3.3 Perception distance

Social attachment is implemented in SOLACE by the affective perception distance ($PD_{Bond}$). This acts as a spatial filter in sensing nearby agents. Familiarity with attachment figures (strength of social distance bond, $SD_{bond}$, 10 point rating scale) tends to facilitate faster perception (detection and recognition), thereby effectively boosting the normal perception distance ($PD_{Normal}$) range (i.e. familiar persons can be recognized even when far away). This is formalised below by equation 1.

$$PD_{Bond} = PD_{Normal}^{k} * \left(1 + \frac{1}{10} * SD_{Bond}\right). \quad (1)$$

A constant *k* defines the bias imposed by the environment (i.e. visibility, audibility). For full visibility during daytime, k = 1; in low visibility during night time, k = 0.2 and in reduced visibility due to fog/snow, k = 0.8.

Equation (1) produces the following graph for perception distance influenced by bond strength (Fig. 3). Data used include 50 meters [23] as the reference perception distance ($PD_{Normal}$), and the emotional bond strength data of Suviletho et al, 2015 for the French case. The calculated perception distance for a partner at night is 4.1 meters, which is consistent with the findings in [24] for comfortable fixation distance during night time.

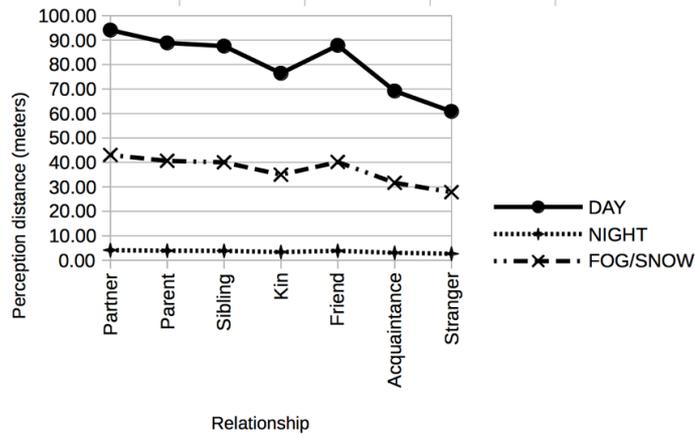

Figure 3: Perception distance influenced by social bond strength.

### 3.4 Physical model environment and UML diagram

The links between the main elements of the model are shown in Fig. 4. There are 5 types of physical environment objects: (1) Buildings, which have different typologies defined by their construction material, use, and height. When buildings are damaged they can produce debris inside a danger zone, which is the area around the building. (2) Soil, the type of which has a huge effect on the seismic intensity; (3) Safe areas, which can be designated shelters, open areas or empty road sections, (4) The earthquake, which can occur at a particular magnitude, duration and intensity, and (5) The individual human agents, with different characteristics and different possible behaviours induced by the earthquake (e.g. move to safe area, seek family members, follow a leader, etc.).

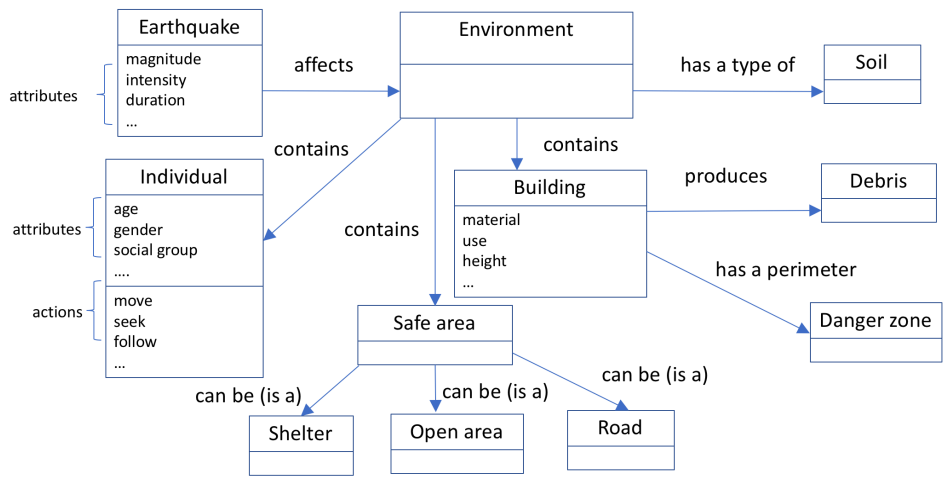

Figure 4: Reduced UML diagram of SOLACE

### 3.5 Decision, behaviours and BDI

A probabilistic approach is applied in choosing pre-evacuation behaviour. During pedestrian evacuation, agent decision-making, behaviour and social interaction are implemented using the BDI approach. Table 2 illustrates several examples of an individual agent's beliefs, desires and possible actions in different contexts. Please note that desires can be contradictory with each other, for instance seeking a family member involves putting oneself in danger. The agent will select the highest priority desire to become its intention, and choose a plan of action to achieve that selected intention.

Table 2: Examples of BDI implemented with social attachment

| CONTEXT | BELIEFS | DESIRES | ACTIONS |
|---|---|---|---|
| Normal situation | None | None | None |
| During extreme earthquake | I'm not safe | Be safe | Seek attachment figures/objects, protect self, egress, evacuate |
| During moderate earthquake | I'm safe, my building is safe | None | None |
| At safe area with family | I'm safe | Stay safe | Stay |
| At safe area but missing family member | I'm safe, my family member is unsafe | Stay safe, family member is safe | Seek family member, call, return to danger area |

### 3.6 Simulation environment

The SOLACE 3D GIS simulation environment and graphical user interface (GUI), developed using the GAMA platform, is shown in Fig. 5. Parameters can be changed in the GUI. Spatial layers (attributes, shape) can be modified with a GIS tool and called directly by the model. It is also possible to change the probability distributions for assigning attributes to model elements. Fig. 5 displays a map of the crisis scenario showing how social attachment drives social interaction and navigation of the synthetic human agents. The chart on the right hand side of the figure shows the progress of evacuation at every simulation time-step. Simulation outputs are saved directly to text (*.csv) files and may be further analysed.

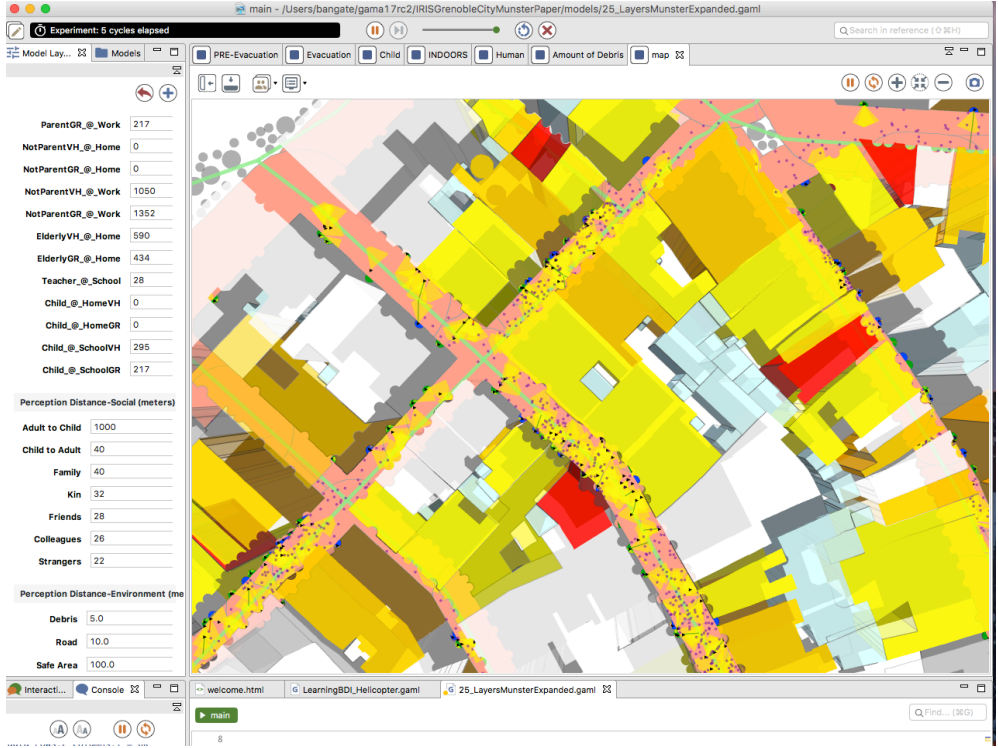

Figure 5: SOLACE interface showing agent social interaction in the crisis environment

4 EXPERIMENTS

The study area is Grenoble, French Alps, an area with moderate seismicity. Earthquakes are rare, which may have led to the population having a low awareness of seismic risk [25]. A large historic sector lies at city centre that contains many old structures. Two districts, Grenette and Crequi-Victor Hugo, in the historic centre were chosen for the experiment. The geographic data used include (a) buildings from IGN BDTOPO(c) [26], and (b) roads and points of interests from OpenStreetMap [27]. Vulnerability and damage probabilities assigned to buildings are from [28]. Population related data from the national census [30].

Table 3: Scenarios tested

|  | Time | Intensity | Disabled agents included |
|---|---|---|---|
| Scenario 1 | Day | 6 | No |
| Scenario 2 | Night | 6 | No |
| Scenario 3 | Day | 6 | Yes |
| Scenario 4 | Day | 8 | No |

Four earthquake scenarios were tested (Table 3). The parameters for the human agents in the model are: (a) perception distance as in Fig. 3.; (b) agent evacuation speeds (maximum velocity in m/sec) from [5]; (c) population of different age groups in the study area: child (0-2 and 3-14), adult (15-29 and 30-59) and elderly (60+). Disabled agents are present in all age categories, except for children since no data was available in [29]. The percentages in Table 4 are used as probabilities. The distribution of agents (absolute numbers) at different times and locations in Table 4 are derived from [30]. Depending on the scenario, agents are initially generated indoors at home, work, school or outdoors. Parent and teacher roles are assigned to adults in the 30-59 age group.

Table 4: Population distribution at initialization derived from national census statistics [30]

| Age group | % disabled | Speed (m/s) | Number per Location (day/night) | | | | |
|---|---|---|---|---|---|---|---|
| | | | Home | Work | School | Public | Outdoors |
| Child (0-2) | 0 | 0 | 75/83 | 0/0 | 0/0 | 0/0 | 8/0 |
| Child (3-14) | 0 | 0 - 2.23 | 0/331 | 0/0 | 298/0 | 0/0 | 33/0 |
| Adult (15-29) | 1.2 - 2.8 | 0 - 3.83 | 209/1842 | 547/0 | 902/0 | 0/0 | 184/0 |
| Adult (30-59) | 1.3 - 12.3 | 0 - 3.83 | 0/1243 | 1119/0 | 0/0 | 0/0 | 124/0 |
| Elderly (60+) | 10.2 - 36.1 | 0 - 1.11 | 553/853 | 215/0 | 0/0 | 0/0 | 85/0 |

## 5 RESULTS

The simulation results are shown in Figure 6. Graphs show arrivals (in % of the population for each category) of the adult, elderly, child, and disabled agents over 1000 seconds (16min 40sec). The short duration in observation time (in seconds) seeks to capture effects of pre-evacuation behaviours, and evacuations just after a sudden earthquake. This is analogous to the 90 second requirement for commercial aircraft evacuations [31] and escape time (RSET) during building fires [32]. Graphs a and b in figure 6 show the trends for scenarios 1, 2 and 4 respectively for able individuals. Overall arrivals in scenario 1 are greater than scenario 2, followed by scenario 4. The night evacuation in scenario 2 with agents having shorter perception distances may have caused this difference. The fewer number of arrivals in scenario 4 can be attributed to the greater amount of debris from the intensity 8 earthquake, trapping or delaying agents. In Figure 6-c, there are few child arrivals because of the long distance to safe areas, and because of the long distance to safe areas, and because of the extra time that it takes for pre-evacuation behaviours such as teachers trying to search for, and group together the children. The terraced appearance of the graphs indicates long delays in the arrivals of child agents. Figure 6-d, shows results for scenario 3, where arrivals of able-bodied agents are greater than for disabled agents.

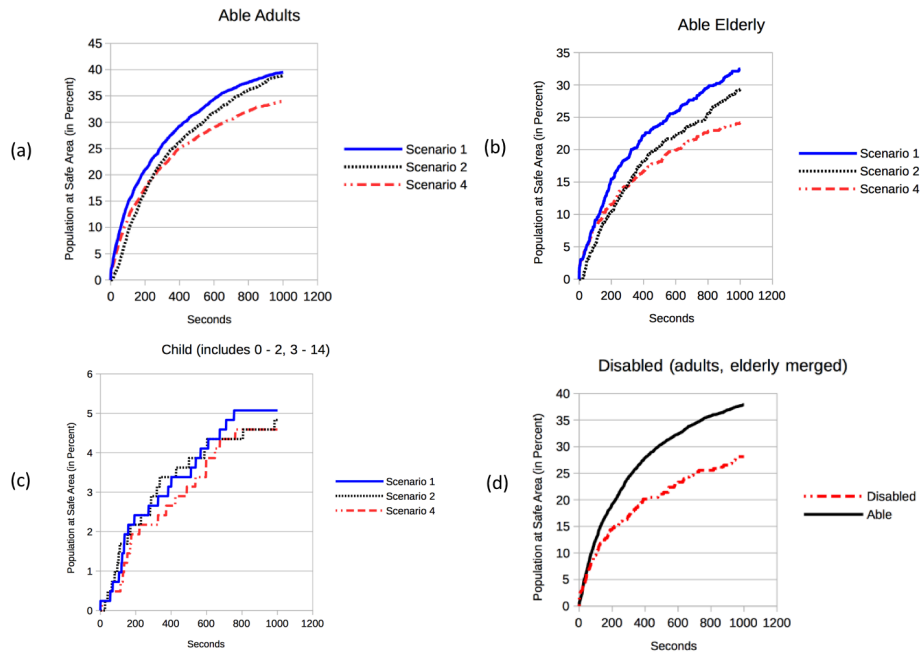

Figure 6: Simulation results

It can be seen that factors that influence arrivals in safe areas include (a) human factors (age, disability, speed), (b) pre-evacuation behaviours, (c) perception distance (social attachment, time of day), (d) social interaction during evacuation, and (e) physical and spatial aspects, such as limitations imposed by debris, and the distance to safe areas.

## 6 DISCUSSION AND CONCLUSION

The complexity of crisis situations provides a big challenge for modellers to make simulations realistic. It entails the recreation of time sensitive interactions of dynamic physical and social aspects that define vulnerability in real geographic space. Data from many sources (national, census, and building data) were useful in model development. These include total population counts or descriptions, presented as percentages. Spatial data on buildings was available with height information. Characterization of buildings used a probabilistic approach to assign attributes, e.g. assigning typology, vulnerability and damage probabilities from [28]. Although qualitative data were available, the challenge was in scaling the data, in percentages, to the level of discrete individual units (e.g. human agents, buildings). A weighted-probabilistic approach was used. Reported populations defined limits and percentages were used as probabilities. This method is promising and can be used to scale datasets (especially field data from surveys) for use in ABM. The experiments showed that many factors must be accounted for to simulate crisis evacuations. Human factors and behaviour are particularly important and the decision to evacuate is critical. Delays in the initial evacuation of agents are governed by pre-evacuation behaviours [33]. A probabilistic approach was used to generate the pre-evacuation behaviour choices and establish an evacuation time delay for each agent. The choice and number of actions undertaken by an agent can delay evacuation. The time of day influences the distribution of people in the city, e.g. at night time most people will be at home. Social attachment bonds typically mean more social interaction between agents, which may delay evacuation, e.g. parents trying to find their children. Similar studies have also shown that social attachment normally results in evacuation delays. The following limitations of the work will be investigated in future: (a) using real mobility data to account for the increase in day time working population, (b) using survey results to define attachment bonds and probabilities for pre-evacuation behaviour, (c) increasing the spatial scale to cover the entire city. However this would also increase the number of agent interactions and would require the use of a high-performance computing (HPC) infrastructure (grid or cluster). To conclude, the model integrates social attachment and spatial elements of evacuation behaviour in simulating seismic crisis. Although numerous data are included in the model, more calibration from survey data is needed. This research finds that integrating both social and physical aspects lead to a more realistic representation of vulnerability in risk analysis.

ACKNOWLEDGEMENTS

This work acknowledges funding from ARC7 (Région Rhône-Alpes) and IXXI. Stéphane Cartier, Philippe Guéguen, Christelle Salameh, Cecile Cornou, Sonia Chardonnel, Bertrand Guillier and Mahfoud Boudis provided invaluable inputs.